\RequirePackage{rotating}
\documentclass[fleqn,usenatbib]{mnras}

\usepackage{newtxtext,newtxmath}

\usepackage[T1]{fontenc}
\usepackage{ae,aecompl}

\usepackage{graphicx}	
\usepackage{amsmath}	
\usepackage{amsfonts}
\usepackage{caption}
\usepackage{pdflscape} 
\usepackage{comment}
\usepackage[dvipsnames]{xcolor}
\usepackage[normalem]{ulem}
\usepackage{appendix}
\usepackage{rotating}
\usepackage{tabularx}
\usepackage{booktabs}

\newcommand{\kms}{km~s$^{-1}$}

\newcommand{\unitlum}{erg~s$^{-1}~$}

\newcommand{\ha}{H$\upalpha$}
\newcommand{\hb}{H$\upbeta$}

\newcommand{\oiii}{[\ion{O}{iii}]}

\newcommand{\chisq}{$\rm \chi^2$}
\newcommand{\css}{CSS100217}

\title[The fall of CSS100217]{The fall of CSS100217: a tidal disruption-induced low state in an apparently hostless active galactic nucleus}

\author[G. Cannizzaro et al.]{
G. Cannizzaro$^{1,2}$ \thanks{E-mail: g.cannizzaro@sron.nl},
A. J. Levan$^{1}$,
S. van Velzen$^{3}$,
G. Brown$^{4}$
\\
$^{1}$Department of Astrophysics/IMAPP, Radboud University, P.O. Box 9010, 6500 GL Nijmegen, the Netherlands\\
$^{2}$SRON, Netherlands Institute for Space Research, Sorbonnelaan, 2, NL-3584CA Utrecht, the Netherlands\\
$^{3}$Leiden Observatory, Leiden University, PO Box 9513, 2300 RA Leiden, The Netherlands\\
$^{4}$Royal Observatory Greenwich, Blackheath Ave, London SE10 8XJ, UK
}
\date{Accepted XXX. Received YYY; in original form ZZZ}

\pubyear{2021}

\begin{document}
\label{firstpage}
\pagerange{\pageref{firstpage}--\pageref{lastpage}}
\maketitle

\begin{abstract}
\css\ was a nuclear, rapid and luminous flare in a Narrow-Line Seyfert 1 galaxy, whose initial interpretation as a supernova is now debated between variability of the active galactic nucleus (AGN) or a tidal disruption event (TDE). In this paper, we present and discuss new evidence in favour of a TDE or extreme flaring episode scenario. After the decay of the flare, the galaxy entered a long-term low luminosity state, 0.4 magnitudes lower than the pre-outburst emission in the V band. We attribute this to the creation of a cavity in the accretion disk after the tidal disruption of a star in a retrograde orbit with respect to the accretion disk rotation, making a TDE our favoured interpretation of the flare. 
We also show how the host galaxy shows a point-like, compact profile, no evidence for an extended component and a relatively low mass, unlike what expected from an AGN host galaxy at z=0.147. A compact host galaxy may result in an increased TDE rate, strengthening our interpretation of the event.
\end{abstract}

\begin{keywords}
Accretion:accretion discs -- transients:tidal disruption events -- galaxies:nuclei -- galaxies: active -- galaxies: peculiar
\end{keywords}

\section{Introduction}
Recent progress in wide-field optical and X-ray surveys have discovered a plethora of rare forms of 
nuclear variability in galaxies. Otherwise quiescent nuclei can be seen to flare on timescales from days to months, commonly attributed to the disruption and subsequent accretion of material from a close approach star -- a so-called tidal disruption event \citep[TDE,][]{Rees1988,Phinney1989}. 
Similarly, but perhaps more mysteriously, a growing number of active galactic nuclei (AGN) show an \textit{extreme} degree of variability \citep[e.g.,][]{graham17}, with ample flares, over timescales that are not compatible with standard accretion disk models \citep{dexter19}. Several mechanisms have been proposed to explain such events: microlensing, a change in obscuration, a supernova (SN) explosion happening in proximity of the AGN, a TDE, or some form of yet to be understood accretion disk variability, such as cooling/heating fronts travelling through the disk \citep{Noda2018} or magnetically launched winds that trigger an increase in the accretion rate \citep{cannizzaro20}.

One such flare is CSS100217:102913+404220 (\css\ hereafter, \citealt{drake11}), an outburst in a known narrow-line Seyfert 1 galaxy (NLS1) at a redshift z=0.147, which translates to a luminosity distance of D=703 Mpc, assuming a cosmology with $H_{\rm 0}$=67.7 km s$^{-1}$ Mpc$^{-1}$, $\Omega_{\rm M}$=0.309, $\Omega_{\rm \Lambda}$=0.691 \citep{Planck2014a}.
The nucleus of the galaxy brightened by over 1 magnitude with respect to the baseline emission, over around 50 days. In \citet{drake11}, the authors discussed three possible scenarios to explain the observed flare: a TDE, AGN variability or a type IIn SN, ultimately favouring the latter interpretation. A TDE was excluded on the basis of \css\ not following theoretical predictions (e.g., blackbody temperature, lightcurve decay rate and peak luminosity) and an AGN outburst was deemed an unlikely explanation because the variability observed in \css\ was an outlier when compared with other Seyfert galaxies, which only exhibit such high amplitude variability on much longer timescales. Additionally, the authors measured variations in the narrow components of the Balmer lines which are not to be expected in an AGN on the timescale of the observed outburst, due to the size of the narrow line region.

In the 10 years after the publication of \citet{drake11}, the number of observed TDEs has increased dramatically, and their observational properties, especially in the optical, show a great degree of heterogeneity \citep[see][for a review]{vanvelzen20review} in terms, for example, of the decay rate and peak luminosity. In addition, the current TDE sample comprises mostly TDEs arising in inactive galaxies and therefore the parameter space of TDEs occurring around already accreting SMBH is under-explored. The exclusion of a TDE as a possible explanation for \css\ appears now less robust. Indeed, already in \citet{blanchard2017}, the authors suggested that \css\ could be explained by a TDE, due to the similarities with PS16dtm, another nuclear transient in an AGN and the primary object of their investigation.
Similarly, the number of AGNs showing strong flux - and, sometimes, spectral - variability is growing \citep[e.g.][]{lamassa15,lawrence16,rumbaugh18} and the variability observed in \css\ is now less of an outlier when compared with other ``hypervariable'' AGNs. Indeed, already \citet{moriya2017} proposed that in \css\ (and PS16dtm) the observed variability was due to interaction between disk winds and the broad line region (BLR) of the AGN. \css\ shares similarities with the five nuclear flares presented in \citet{frederick2021}, all happening in NLS1 galaxies - a type of host galaxy which seems preferential for such events. Interestingly, the authors underline how such flares in NLS1 galaxies can be mistaken for SN IIn explosions, while they interpret them as either AGN activity or a TDE.
Finally, a strong reason for the present re-discussion of the nature of the flare is the long term behaviour of the lightcurve: after the flare decayed, the galaxy entered a lower emission state, around 0.4 mag lower than the pre-outburst level of emission (Figure~\ref{fig:lcmag}). This low state continues up to at least the beginning of 2021, as suggested by measurements of different all-sky surveys. This low state was mentioned in \citet{moriya2017}, but was not a determining factor in their interpretation of the event.
The coincidence between the decay of the flare and the onset of the low state suggests that the two phenomena are related and, being in the presence of an AGN, it is straightforward to connect the observed behaviour to the accreting SMBH, as it is difficult to envision how a nuclear SN explosion could be at the origin of the $\sim$10 years long low emission state.

\section{Data and data reduction}

\subsection{Optical spectra}
We retrieve three optical spectra of the source obtained before, at the peak and after the transient event. The pre-outburst spectrum was obtained during the legacy survey of the the Sloan Digital Sky Survey \citep[SDSS,][]{sdss14}, on 2003 December 29. The spectrum at peak was obtained on 2010 May 18, with the Low Resolution Imaging Spectrometer \citep[LRIS,][]{lris-keck}, mounted at the Cassegrain focus of the Keck I telescope, with grism 400/3400 for the blue arm and grating 400/8500 for the red arm, with nominal resolution $\Delta\lambda$ 6.8 \AA\ and 6.9 \AA for a 1.0'' slit, respectively. We also obtained a late post-outburst spectrum on 2016 April 26, with the Auxiliary-port CAMera (ACAM) spectrograph, mounted at the Cassegrain focus of the William Herschel Telescope (WHT), in combination with the V400 grism with a nominal resolving power R=430 at 5650 \AA\ with a 1.0'' slit. The LRIS and ACAM spectra were reduced with standard {\sc iraf} \citep{tody86} procedures such as flat field and bias correction, cosmic ray removal \citep{lacosmic} and wavelength and flux calibration with arc lamps and standard star observations. In the case of the LRIS spectrum, the \hb\ and \oiii\ emission lines fall in the area of overlap between the two arms, and therefore the flux calibration may not be perfect (being this wavelength region at the right/left edge of the blue/red arm, respectively). The resulting spectroscopy is shown in Figure~\ref{fig:optspec}. 

\subsection{Optical and UV photometry}

    \begin{figure*}
        \includegraphics[width=1.\textwidth]{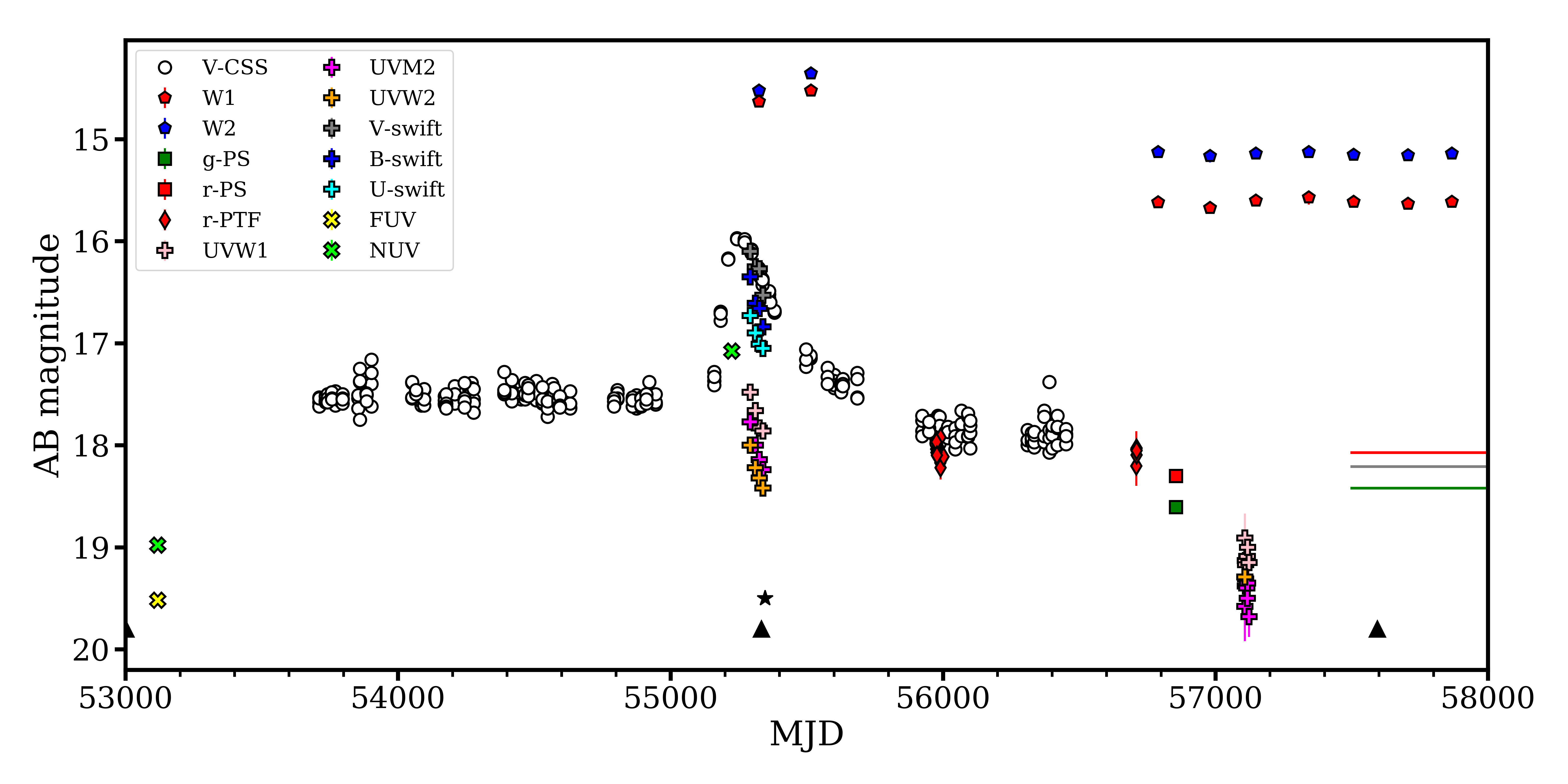} 
        \caption{Light curve of \css, in AB apparent magnitudes versus MJD. The different filter/instruments are indicated in the legend. The red and green solid lines on the right hand side of the plot indicate the mean value of the magnitude of the object in the r and g filters respectively, obtained from ZTF. The grey solid line is the value in the V filter, obtained from the latter two through the filter transformations of \citet{jordi06}. The three black triangles at the bottom of the plot indicate the three epochs of optical spectroscopy and the black star indicates the epoch of the HST images.}
        \label{fig:lcmag}
    \end{figure*}

We include optical and UV photometric measurements of the source, obtained by different instruments and surveys: the Catalina Sky Survey (CSS, \citealt{Drake09}), Pan-STARSS (PS, \citealt{panstarrsdata}), the Zwicky Transient Facility (ZTF, \citealt{ztf}), the Palomar Transient Factory (PTF, \citealt{ptf}), the Galaxy Evolution Explorer (GALEX, \citealt{galex}) and the Neil Gehrels \textit{Swift} observatory \citep{swift:gehrels}. For the \textit{Swift} data, magnitudes (see Table~\ref{tab:UVOTobs}) were obtained using a 5 arcsec aperture centred on the source and a 20 arcsec aperture centred on an empty nearby region to estimate sky background levels. The rest of the magnitude values are provided by the respective telescope/instrument institution.

\css\ has been observed once during the outburst, on 2010 May 31, with the Wide Field Camera 3 (WFC3) on the Hubble Space Telescope (HST), obtaining images in the F390W, F555W and F763M filters. We retrieved the images from the HST archive\footnote{https://mast.stsci.edu/} and reduced them via the {\sc astrodrizzle} package to obtain images with a 0.025 arcsecond pixel$^{-1}$ plate scale. Details of the HST observations are listed in Table~\ref{tab:XRT-HSTobs}.

\subsection{{\it\bf Swift}/XRT}
The source was observed in the X-rays with the XRT instrument on board the {\it Swift} satellite several times, both during the outburst and after its end. The data were reduced and analysed with the online {\it Swift}/XRT pipeline tool\footnote{\url{https://www.swift.ac.uk/user\_objects/}} \citep{evans09}, that uses {\sc heasoft} v6.28. The resulting lightcurve is shown in Figure~\ref{fig:xrayLC}. Due to the low number of counts, we stack all available spectra to create a single spectrum, with total exposure time of around 22 ks. Details of the {\it Swift} observations are listed in Table~\ref{tab:XRT-HSTobs}.

\subsection{WISE}

The field of \css\ has been observed several times by the Near-Earth Objects Wide Field Survey Explorer (NEO-WISE, \citealt{mainzer11}) in the W1 and W2 bands. We retrieve the individual epoch photometry from the ALLWISE data release and subsequent NEOWISE re-activation\footnote{\url{https://irsa.ipac.caltech.edu/Missions/wise.html}}. A source outburst is clearly detected in the WISE data (see Fig.~\ref{fig:lcmag}). Notably it rises between the first and second epochs even though the first epoch takes place 80 days after the optical peak. This is a suggestion of re-processing of the UV/optical flare into the IR via dust or material surrounding the SMBH.

\begin{sidewaystable}
    \centering
 	\caption{UVOT apparent magnitudes.}

 	\label{tab:UVOTobs}
 	\tabcolsep=0.11cm

	\begin{tabular}{lcccccc}
	    \hline
		MJD$^{(1)}$  & UVW1  & UVM2 & UVW2 & U & B & V\\
        days & mag & mag & mag & mag & mag & mag  \\
        \hline
        55292.6 & 	17.48 $\pm$	0.03 & 	17.77 $\pm$	0.03 & 	18.00 $\pm$	0.03 & 16.73 $\pm$	0.04 & 16.35 $\pm$	0.04 & 16.10 $\pm$	0.05\\
        55311.7 & 	17.66 $\pm$	0.04 & 	18.00 $\pm$	0.04 & 	18.22 $\pm$	0.03 & 16.90 $\pm$	0.04 & 16.61 $\pm$	0.04 & 16.25 $\pm$	0.05\\
        55326.1 & 	17.83 $\pm$	0.04 & 	18.14 $\pm$	0.04 & 	18.32 $\pm$	0.04 & 17.01 $\pm$	0.04 & 16.66 $\pm$	0.04 & 16.27 $\pm$	0.05\\
        55339.2 & 	17.86 $\pm$	0.04 & 	18.24 $\pm$	0.04 & 	18.42 $\pm$	0.03 & 17.05 $\pm$	0.04 & 16.84 $\pm$	0.04 & 16.53 $\pm$	0.06\\
        57107.6 & 	18.91 $\pm$	0.24 & 	19.58 $\pm$	0.34 & 	19.29 $\pm$	0.21 \\
        57110.0 & 	19.13 $\pm$	0.11 & 	19.38 $\pm$	0.16 & 	\\	  
        57111.0 & 	19.17 $\pm$	0.07 & 	19.31 $\pm$	0.11 & 	\\	 
        57114.7 & 	19.09 $\pm$	0.18 & 	19.40 $\pm$	0.24 & 	\\	  
        57116.6 & 	19.09 $\pm$	0.10 & 	19.50 $\pm$	0.17 & 	\\	 
        57118.1 & 	19.00 $\pm$	0.10 & 	19.35 $\pm$	0.14 & \\		 
        57123.6 & 	19.15 $\pm$	0.12 & 	19.68 $\pm$	0.20 & \\		  
		\hline
    \end{tabular}

\textit{Note.}(1) Modified Julian Day of observations. All values have been corrected for Galactic extinction.\\
Magnitudes are in the native systems: AB for the UV filters, Vega for the Johnson filters.
\end{sidewaystable}

\begin{table}
	\centering
	\begin{center}
	\small
 	\caption{XRT and HST observations.}

 	\label{tab:XRT-HSTobs}

	\begin{tabular}{lcc}
	\hline
	    \multicolumn{3}{c}{{\em Swift}-XRT observations}\\
		MJD$^{(1)}$  & exposure time  & countrate\\
         days & s & $10^{-3}$cts $\rm s^{-1}$ \\
        \hline
        55292.58 & 3518 & $6.6${\raisebox{0.5ex}{\tiny$^{+2.3}_{-1.9}$}} \\
        55311.69 & 3044	& $2.0${\raisebox{0.5ex}{\tiny$^{+	 1.2}_{	 -0.9}$}} \\
        55326.01 & 3311	& $2.2${\raisebox{0.5ex}{\tiny$^{+	 1.2}_{	 -0.9}$}} \\
        55339.19 & 4027 & $0.3${\raisebox{0.5ex}{\tiny$^{+	 0.6}_{	 -0.3}$}} \\
        57107.63 & 178	& $6${\raisebox{0.5ex}{\tiny$^{+  	 10}_{ -5}$}} \\
        57109.96 & 607	& $37.1${\raisebox{0.5ex}{\tiny$^{+	 8.7}_{	 -8.7}$}} \\
        57110.96 & 1454	& $36.3${\raisebox{0.5ex}{\tiny$^{+	 5.5}_{	 -5.5}$}} \\
        57114.68 & 291	& $3.7${\raisebox{0.5ex}{\tiny$^{+	 6.4}_{	 -3.1}$}} \\
        57116.58 & 704	& $15.1${\raisebox{0.5ex}{\tiny$^{+	 5.9}_{	 -4.7}$}} \\
        57118.14 & 887	& $4.8${\raisebox{0.5ex}{\tiny$^{+	 3.4}_{	 -2.4}$}} \\
        57123.56 & 576	& $12.0${\raisebox{0.5ex}{\tiny$^{+	 6.0}_{	 -4.6}$}} \\
        58860.44 & 767	& $10.6${\raisebox{0.5ex}{\tiny$^{+	 4.8}_{	 -3.7}$}} \\
        58908.11 & 1652 & $2.9${\raisebox{0.5ex}{\tiny$^{+	 2.9}_{	 -1.8}$}} \\
        59304.73 & 789  & $30.1${\raisebox{0.5ex}{\tiny$^{+	 6.8}_{	 -6.8}$}} \\
		\hline
		\hline
		\multicolumn{3}{c}{HST observations}\\
		MJD & Filter & exposure time \\
		   days   &  & s   \\
		\hline
		55347.3 & F390W & 315  \\
		55347.3 & F555W & 245  \\
		55347.3 & F763M & 525  \\
		\hline
        \end{tabular}

\textit{Note.}(1) Modified Julian Day of observations. 
\end{center}
\end{table}

\section{Data analysis and results}

    \begin{figure*}
        \includegraphics[width=1.\textwidth]{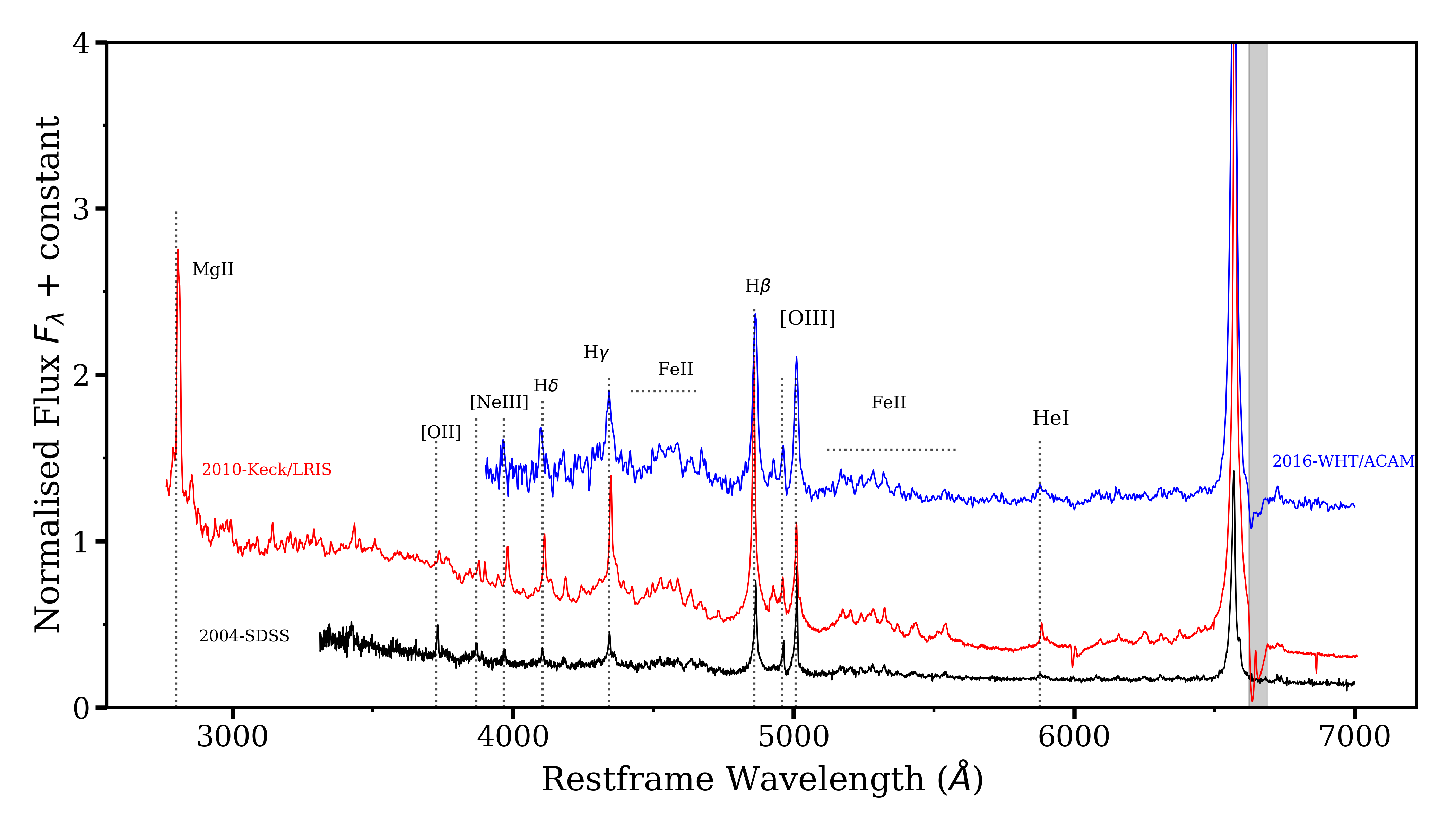} 
        \caption{The three optical spectra of our follow-up campaign are plotted here: the SDSS spectrum from 2004 in black (bottom), the Keck-LRIS spectrum from 2010 in red (middle) and the WHT-ACAM spectrum from 2016 in blue (top). With the grey band, we indicate a region strongly affected by atmospheric telluric absorption. The most prominent emission lines are indicated in the plot. The spectra have been normalised for plotting purposes.}
        \label{fig:optspec}
    \end{figure*}

\subsection{Black Hole Mass estimation}
The optical spectra do not show the Ca H+K doublet (in itself a constraint on the host galaxy), meaning that we cannot employ the $M-\sigma$ relation to estimate the mass of the SMBH. We, instead, use two single-epoch mass measurements based on the continuum luminosity and the flux and width of the \hb\ broad emission line \citep{vestergaard06}. Using the values of the line fit performed on the SDSS, pre-outburst spectrum, we obtain a black hole mass $M_{BH}=(5.7\pm2.7)\times10^7\,M_\odot$ and $M_{BH}=(2.4\pm2.7)\times10^7$, where the uncertainty is dominated by the scatter in the relation. We stress that such single-epoch measurements are more suited for studies of statistical samples, and that for a single object these methods are less reliable, yielding only order of magnitude estimates. From here on, we will consider a value of $10^7\,M_\odot$ for the central black hole.

\subsection{Optical spectroscopy}
The sequence of optical spectra is shown in Fig.~\ref{fig:optspec}. These are optical spectra of the galaxy (no subtraction performed) taken before (SDSS), during (Keck/LRIS) and after the outburst (WHT/ACAM).  Qualitatively, the spectra show narrow emission lines of both permitted and forbidden transitions and broad emission lines of permitted transitions, as expected given the NLS1 classification of the galaxy \citep{drake11}. Some of the forbidden lines (e.g. the \oiii\ doublet at 4959, 5007\AA) show a broad base, but this component is detected only for the lines with higher SNR and better contrast with respect to the continuum. The \ha\ line is contaminated by a telluric absorption band. The spectrum taken during the outburst shows a slightly enhanced blue continuum and stronger emission lines. Notably, we do not detect stellar absorption lines.
  
We fit the emission lines with a series of Gaussian curves and a polynomial (of first or second degree, depending on the spectrum) to fit the local continuum, using the package {\sc lmfit} \citep{lmfit}. During the fitting procedure, we tied together the width of the narrow lines and the wavelength separation of known line doublets. We also masked the wavelength region affected by the telluric absorption band. This may have hindered the results of the fitting procedure for the broad \ha\ component. In the higher resolution spectra, some of the lines with the best SNR show evidence of an intermediate velocity component, as seen also in \citet{drake11}. 
The flux and full width half maximum (FWHM) of both the narrow and broad component of the \ha\ and \hb\ lines are listed in Tab. \ref{tab:linefit}. The values have been corrected for the instrumental broadening, measured from the arc lines\footnote{for the SDSS spectrum, we considered a broadening of $\rm \sim 90$\kms, as reported on the SDSS website.} and corrected for the seeing, when it was lower than the slit width. We stress that the resolution of the last spectrum, obtained with the ACAM spectrograph, is much lower than the resolution of the other two spectra. Given that the lines may all have a broad component, even if not always resolved, the lower resolution causes the narrow component to capture flux from the broad base and therefore the resulting fluxes can be overestimated. The higher value of the flux of the narrow \ha\ and \hb\ lines in the last epoch is probably due to this effect. Similarly, the higher value of the FWHM may be due to a non-perfect correction for the instrumental broadening, as narrow features are inherently difficult to investigate with a low resolution instrument. In the central epoch, the flux of the broad components, especially the \ha, is higher, in accordance with the increase in flux in the optical band during the outburst. The FWHM of the lines does not vary significantly between the three epochs. Although we were not able to retrieve all the spectra analysed in \citet{drake11}, in the three spectra included in this manuscript, we detect changes in the narrow lines, but these variations could be attributed to the different instruments employed. The results of the line fitting procedure are listed in Table~\ref{tab:linefit}.

\begin{table*}
	\centering
	 \begin{center}
	\small
 	\caption{Results from the line fitting procedure.}

 	\label{tab:linefit}
 	\tabcolsep=0.11cm

	\begin{tabular}{lcccccccc}
	\hline
		MJD$^{(1)}$  & $\rm H\upbeta_n$ flux  & $\rm H\upbeta_n$ FWHM & $\rm H\upbeta_b$ flux & $\rm H\upbeta_b$ FWHM   & $\rm H\upalpha_n$ flux  & $\rm H\upalpha_n$ FWHM & $\rm H\upalpha_b$ flux & $\rm H\upalpha_b$ FWHM 	\\
         days & 	$\rm 10^{-16}erg\,cm^{-2}\,s^{-1}$ & $\rm km\,s^{-1}$ & $\rm 10^{-16}erg\,cm^{-2}\,s^{-1}$ & $\rm km\,s^{-1}$ & $\rm 10^{-16}erg\,cm^{-2}\,s^{-1}$ & $\rm km\,s^{-1}$ & $\rm 10^{-16}erg\,cm^{-2}\,s^{-1}$ & $\rm km\,s^{-1}$ \\
        \hline
        53002 & 21.0 $\pm$ 2.7 & 320 $\pm$ 17 & 54 $\pm$ 3.7 & 1675 $\pm$ 71  & 138.3 $\pm$ 7.9 & 528   $\pm$ 16  & 127 $\pm$ 12 & 2604 $\pm$ 91 \\
        55334 & 77.1 $\pm$ 4.4 & 432 $\pm$ 17 & 139  $\pm$ 10  & 1785 $\pm$ 73  & 386.9 $\pm$ 8.0 & 488.1 $\pm$ 7.6 & 701 $\pm$ 19 & 2629 $\pm$ 37 \\
        57504  & 104  $\pm$ 17  & 782 $\pm$ 37 & 86   $\pm$ 33  & 2280 $\pm$ 450 & 540   $\pm$ 14  & 797   $\pm$ 11  & 323 $\pm$ 40 & 2420 $\pm$ 140 \\
		\hline
\end{tabular}

\textit{Note.}(1) Modified Julian Day of observations. The values are corrected for the instrumental broadening (see body of text for details). The subscripts $n$ and $b$ indicate the narrow and broad component of the emission line, respectively.
\end{center}
\end{table*}

\subsection{UV, optical and IR photometry}
\label{sec:UVOptphot}

In Fig.~\ref{fig:lcmag} we plot the optical and UV light curve of \css. The source shows an intense flare, with amplitude $\rm>1.5$ mag, a very steep rise to a peak luminosity of $\rm \sim5\times10^{44}$ \unitlum (in the V band) over around 50 days, and a shallower decay that lasted $\rm \sim 600$ days. Due to the gaps in the CSS light curve, it is difficult to have a precise estimation of the rise and decay time. The outburst is observed also in the UV, by comparing Swift UVOT observations in the UVW1, UVM2 and UVW2 filters with archival GALEX observations in the NUV and FUV bands, performed long before the onset of the optical flare. The V band lightcurve has a decay rate $\propto t^{0.24}$ (we considered the portion of the CSS lightcurve from peak until $\rm MJD\sim52650$).
\citet{drake11} performed an accurate analysis of the energetics of the flare, thus we will use their results in the present discussion of \css. The flare radiated a total energy of around $\rm10^{52}$ erg and the bolometric luminosity at peak is $\rm\sim1.5\times10^{45}$ \unitlum, very close to the Eddington luminosity for a $\rm10^7\,M_\odot$ BH: $\rm1.3\times10^{45}$ \unitlum.
Perhaps the most striking feature of the light curve is the lower level of emission reached at the end of the outburst, compared to the pre-flare luminosity: in fact, the latest V band magnitudes reported by CSS show a drop of $\rm\sim0.4$ mag in baseline emission. We checked nearby sources of similar magnitudes in the CSS archive, to evaluate if the drop in luminosity is due to calibration errors. The lightcurve of these comparison objects does not show a similar drop, meaning that the observed low state is a property of \css.
By comparing the post-flare g and r band magnitudes of PS, PTF and ZTF (Fig. \ref{fig:lcmag}), it appears that the object is still in this low state, up to January 17, 2021 (MJD 59231, the last epochs of ZTF r band data). We also check the  Asteroid Terrestrial-impact Last Alert System (ATLAS) forced photometry \citep{atlasforced} at the position of the galaxy and we see that between April 2016 and May 2021 the galaxy shows a constant luminosity in the ATLAS o and c bands \citep{atlas}. We therefore consider that the galaxy is still in the low emission state reached after the end of the flare.
In Fig.~\ref{fig:lcmag}, we also plot the mid-IR photometry from NEO-WISE, in the W1 and W2 filters. NEO-WISE takes $\sim$10 exposures per epoch of observation and we plot the average magnitude values of these snapshots (data are binned at the ~6 months epoch). While the time-coverage of the NEO-WISE lightcurve has a large gap, it seems that the mid-IR emission does not trace the optical one, but it may arise from reprocessing of the radiation by dust in the system (e.g., the dusty torus).\\
Following \citet{vanvelzen2021neutrino}, we measure the amplitude of the flare as the increase in flux in the first year of WISE data $\rm (\Delta F)$ divided by the RMS of the MIR flux $\rm (F_{rms})$ after this flare, obtaining $\rm \Delta F/F_{rms}=32$ in W1. This relatively large {\it echo strength} is typically only observed for TDEs and extreme AGN flares \citep{vanvelzen2021neutrino}. The post-flare WISE data show low-amplitude, but statistically significantly variability (in the W1 filter the reduced \chisq\ is 36, for a model with a constant flux), which is to be expected since the W1-W2 color is in the AGN region \citep{Stern2012}.

\subsection{X-rays}
\label{sec:xraydata}
    \begin{figure}
        \includegraphics[width=.49\textwidth]{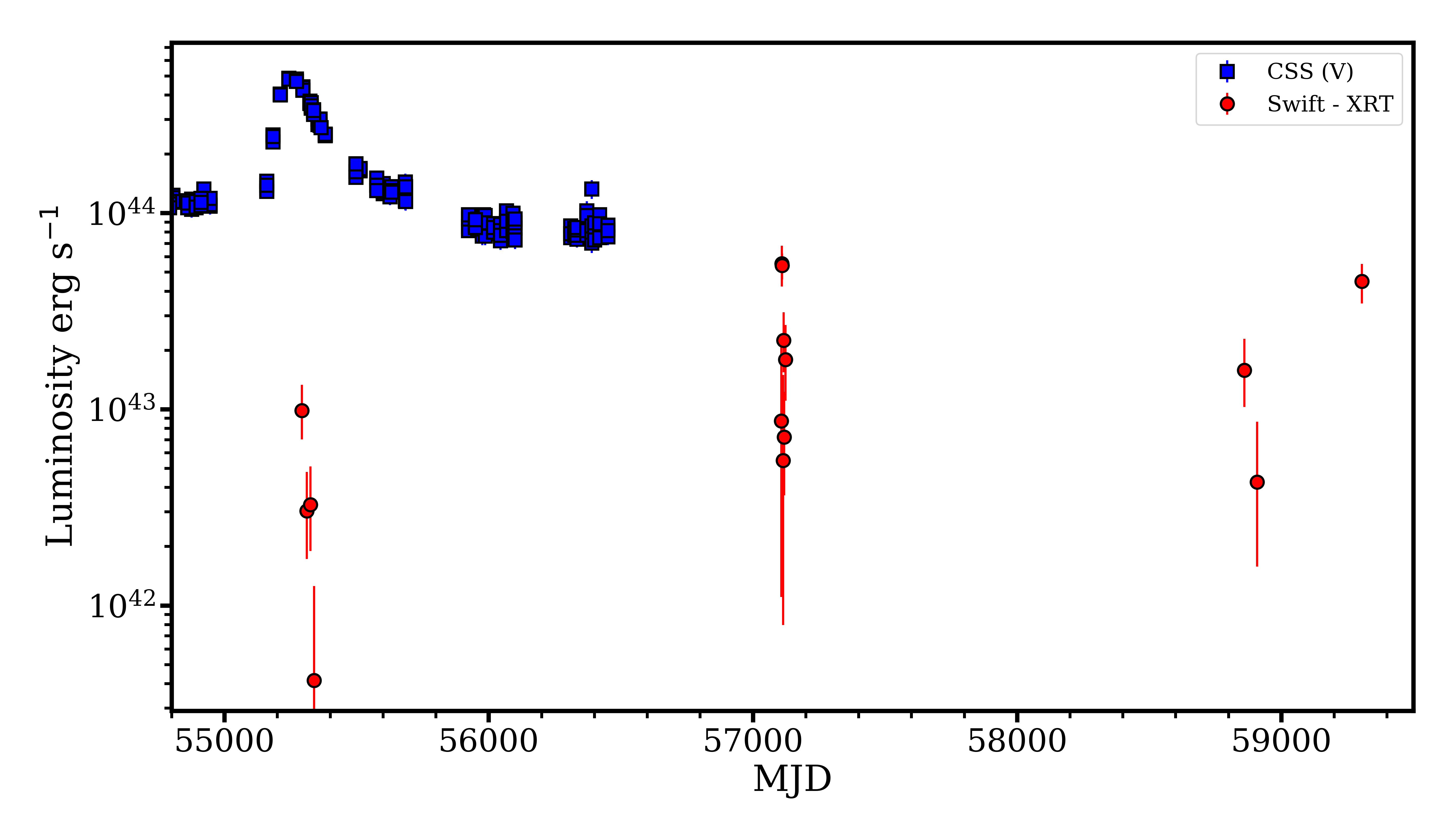}
        \caption{Plotted here are the V band (CSS, blue squares) and 0.3--10 keV (XRT, red circles) luminosities with respect to time (in MJD).}
        \label{fig:xrayLC}
    \end{figure}

Using the online {\it Swift} XRT pipeline, we fit the stacked spectrum with a power-law with photon index $\rm \Gamma=3.3^{+0.8}_{-0.4}$, with Galactic $\rm N_H=1.02\times10^{20}\, cm^{-2}$ \citep{willingale2013}. This results in a counts-to-flux conversion factor of $\rm 2.5\times10^{-11} erg\,cm^{-2}\, ct^{-1}$ and $\rm 3.2\times10^{-11} erg\,cm^{-2}\, ct^{-1}$ for the observed and unabsorbed flux, respectively, in the 0.3--10 keV band.\\
The X-ray 0.3--10 keV light curve is shown in Fig. \ref{fig:xrayLC}, where we also plot the V band luminosity from CSS, for comparison. Interestingly, the X-ray luminosity seems to be somewhat anti-correlated with the optical/UV flare. On top of the clear absence of a flare in the X-ray lightcurve, the post-flare X-ray luminosity is seen increasing during the optical low emission state. Moreover, there is relatively strong intra-day variability observed in all groups of observations. This variability is still present when considering only the XRT observations with enough counts for the resulting spectrum to be fit alone. The flux values obtained from the fit to the single spectra are in good agreement with the corresponding values obtained with the conversion factor noted above. We are not able to check for variations in the spectral slope, as in the single spectra the powerlaw index is very poorly constrained.\\
In light of the observed X-ray variability, we checked for variability in the optical/UV bands, by using the photon-counting properties of the UV Optical Telescope (UVOT) on board the {\it Swift} satellite, using the task {\sc uvotmaghist} to create a lightcurve out of each single exposure. This task produces 2--3 datapoints for each observations, depending on the length of the exposure and the countrate. We do not observe variability on a $\sim$seconds timescale. This implies that the optical/UV observations are not directly tracking the X-ray emission.

    \begin{figure}
    \centering
        \includegraphics[width=.4\textwidth]{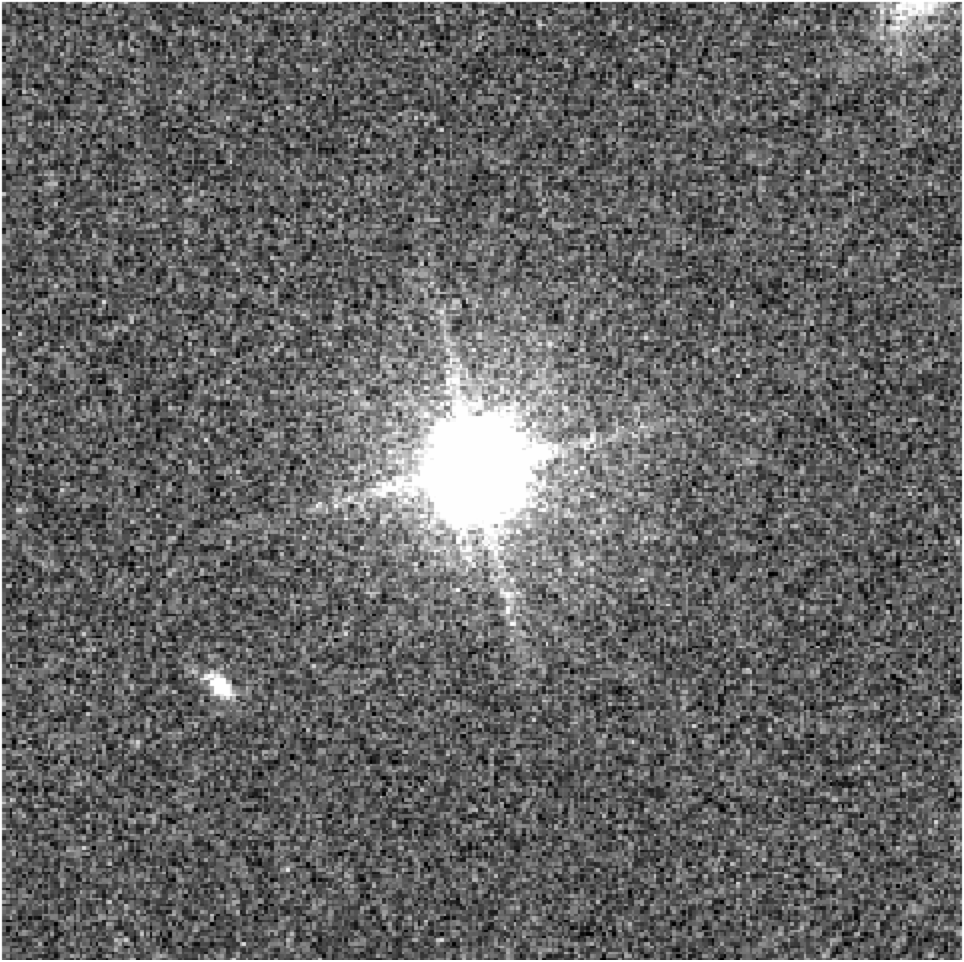}
        \includegraphics[width=.4\textwidth]{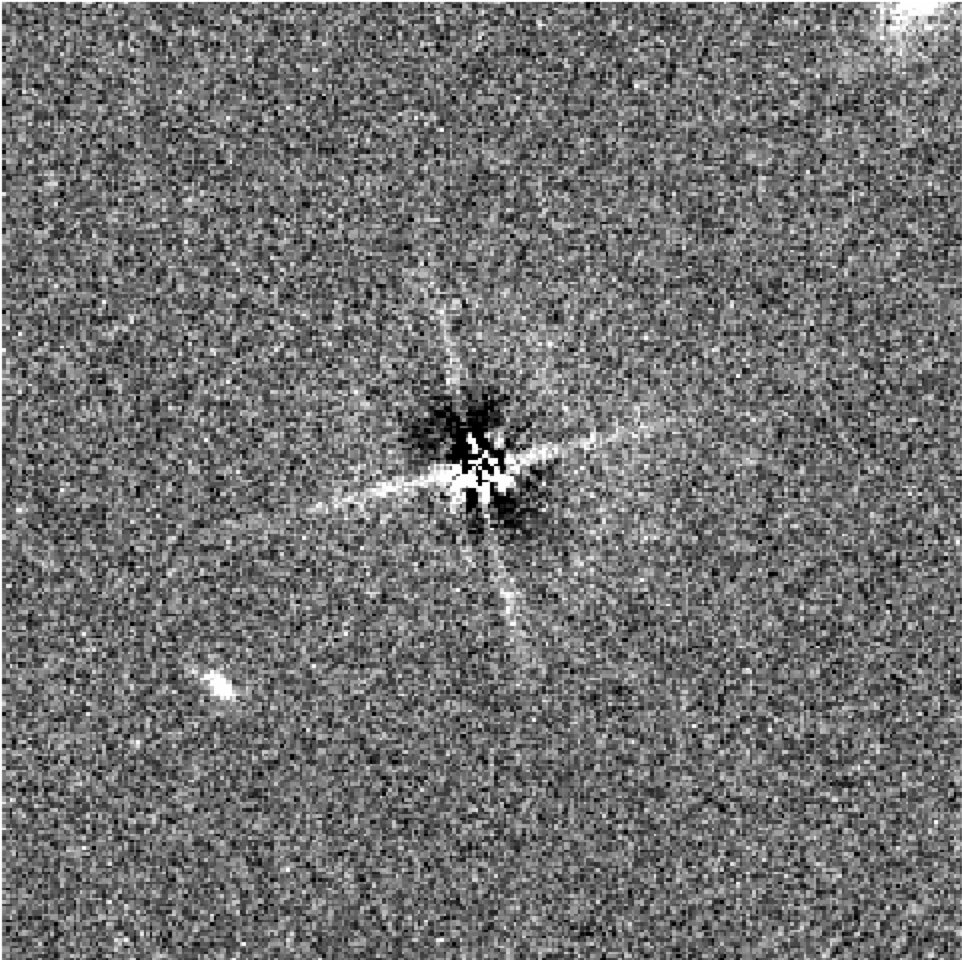}
        \caption{{\it Top}: zoomed in, HST F555W image of \css. {\it Bottom}: The same image after the subtraction of the galaxy model built with \textsc{galfit}. Images are squares of side $\sim$12 arcsec.}
        \label{fig:galfit}
    \end{figure}

   \begin{figure}
        \centering

        \includegraphics[width=\columnwidth]{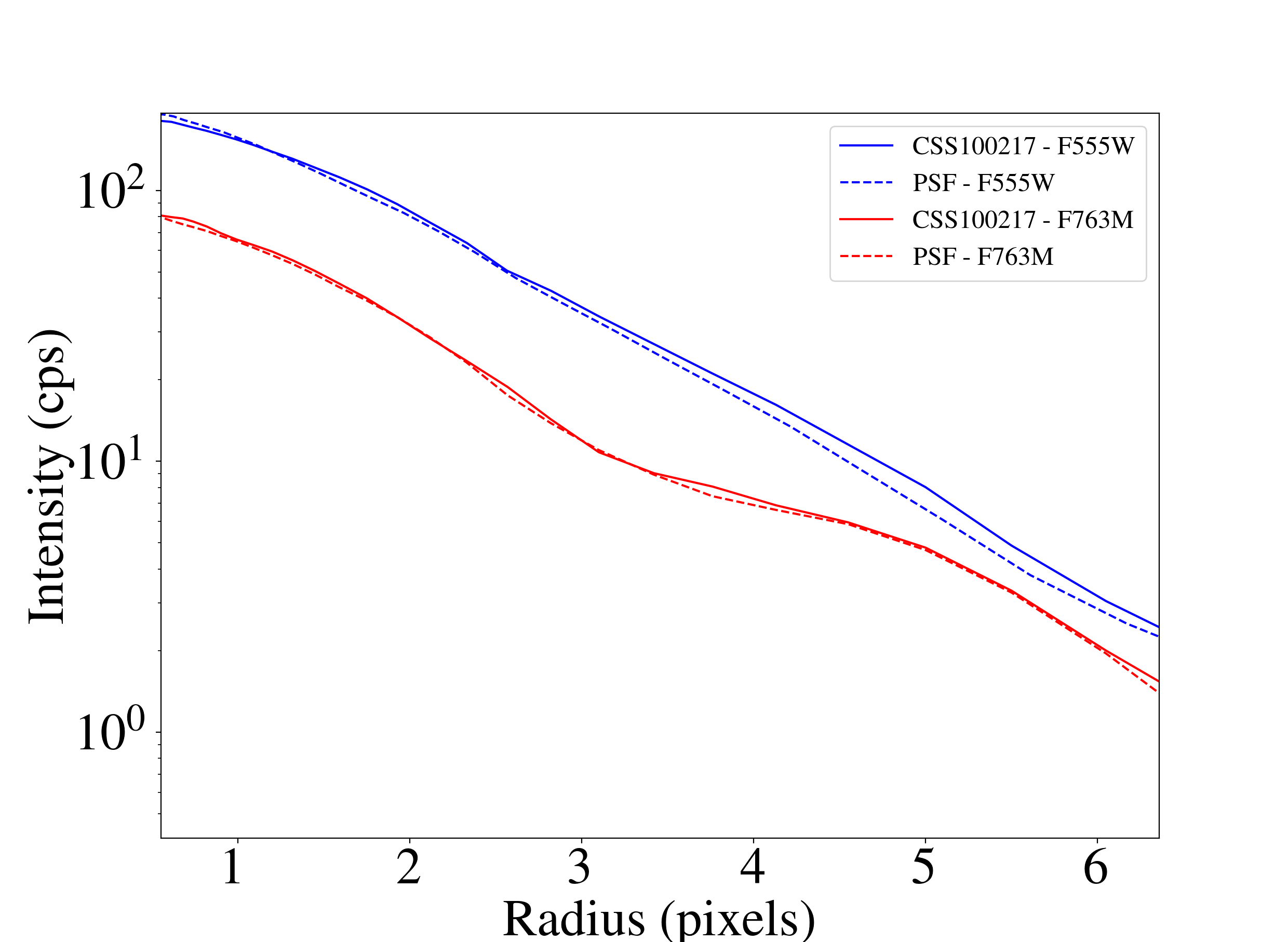}

        \caption{Comparison of the light profiles of \css\ (solid lines) and of a PSF (dashed lines) built from a different HST image, taken with the same setup as our data for the F555W (blue, top) and F763M (red, bottom) filters.}
        \label{fig:PSF_comparison}
    \end{figure}

\subsection{Host galaxy}
\label{sec:galfit}
We investigate the properties of the host galaxy using the high resolution HST images. Although the images are taken close to the peak of the outburst, they are striking in showing an apparently point-like source, with little evidence for any underlying host galaxy. We quantify this by attempting to fit and subtract the galaxy via {\sc galfit} \citep{galfit2002,galfit2010}, and via a direct comparison of the radial profile of the source with point sources in other {\em HST} images obtained in the same filters via isophotal fitting (see Fig.~\ref{fig:galfit},  and Fig.~\ref{fig:PSF_comparison}). In both of these cases, the system is consistent with being dominated by a point-like source, and no evidence of any extended component is found. The implication is that any galaxy is more compact, or comparably compact to the {\em HST} point spread function. At $z=0.147$, this implies a size of less than a few hundred parsecs, comparable to ultra compact dwarfs \citep[e.g.][]{zhangbell17}.

During the {\sc galfit} fitting procedure, we were unable to include a PSF model: The field of view of the HST images does not have other sources from which to build a PSF and using a pre-constructed PSF\footnote{see\url{https://www.stsci.edu/hst/instrumentation/wfc3/data-analysis/psf}, where also a description of the limits of such PSF model is presented.} did not provide an improved fit. Indeed, the central region of the galaxy, after the subtraction, remains noisy (Fig.~\ref{fig:galfit}, bottom panel). We obtain an acceptable fit of the galaxy (reduced \chisq=1.6) using two Sersic models, with Sersic indexes 0.7 and 7.0 and half-light radii 0.05 and 0.15 arcsec, respectively.

We can also use the available IR photometry to place limits on the stellar mass of the galaxy. The  archival K-band magnitude ($\rm m_K=14.55\pm0.07$, from the Two Micron All Sky Survey - 2MASS, \citealt{2mass}) is probably dominated by the AGN component. However, under the assumption that it were dominated by stellar light we can infer a stellar mass $M_{\star,gal}$. Following the method proposed in \citet{longhetti2009}: $M_{\star,gal} < 3\times10^{10}\,M_\odot$. Such a galaxy would typically be expected to have a size of $>1$ kpc  \citep{shen2003}, although the size would also reduce when accounting for likely AGN contamination in the K-band.

\section{discussion}

\subsection{The nature of the flare}

In light of the new discoveries since the first publication discussing the nature of \css\ \citep{drake11}, in terms of new and more varied observational properties of TDEs and of the growing number of extremely variable AGNs and considering the long-term low emission state, we will discuss again the TDE and variable AGN scenarios as viable explanations for the observed \css\ flare.
Additionally, following the analysis of \citet{vanvelzen2021neutrino}, we obtain P(extreme)/P(AGN)=0.007, where P(extreme) indicates the probability of the flare to be a high amplitude accretion flare and P(AGN) that it is due to normal AGN variability\footnote{In \citet{vanvelzen2021neutrino}, P(extreme) was noted as P(TDE), to indicate the probability of the flare to be caused by a TDE. Here we use P(extreme) to avoid confusion between the two scenarios we propose.}, meaning that it is very unlikely that the observed flare is due to regular AGN variability.

\subsubsection{Tidal disruption event}

A tidal disruption event happens when a star gets torn apart by the tidal field of an SMBH, with roughly half of the stellar material being bound to the black hole and accreting onto it, giving rise to a rapid and luminous flare \citep{rees88,Phinney1989}. Observing TDEs happening in an AGN is inherently difficult, due to the ambiguity between AGN variability and a TDE-induced flare. Nonetheless, simulations have shown how the interaction of the tidal stream and the accretion disk can cause the disk to be quickly drained from the point of impact inwards, causing a surge of accreting material onto the SMBH, which translates to the rapid and luminous flare observed (\citealt{chan19}, see also \citealt{cannizzaro20}). The disk will then replenish itself on the viscous timescale \citep[e.g.,][]{dexter19}, which is much longer than the timescale at which the disk is drained, causing the observed low-state. Interestingly, \citet{chan19} speculates that the X-ray emitting corona may be disturbed, causing a dip in the X-ray emission, which would explain the lower X-ray emission at the peak of the flare observed in \css. This TDE is conceptually similar to a ``regular'' TDE (i.e., happening around a dormant SMBH), with a star being scattered towards the SMBH and inside the tidal radius, just with different behaviour of the stream of disrupted material and an overall bigger mass reservoir, depending on how big a portion of the accretion disk gets drained. In a subsequent paper expanding the scope of their simulations, \citet{chan2021} show that hard X-ray (and $\upgamma$-ray) emission can be expected from the interaction between the tidal stream and the disk. While the bulk of the energy is predicted to be emitted above 10 keV - outside the range of {\it Swift}/XRT - we note that we do not detect hard X-ray photons in any of our X-ray observations. Of course, due to the scarce time coverage of our dataset, we may simply have missed this high energy emission. The lower X-ray emission at peak, in this case, could be explained by obscuration of the X-ray emitting region by the disrupted material.
\\

Even more relevant is the case of the tidal disruption of a star embedded in the accretion disk: in fact, \citet{mckernan2021} predicts that a star on a retrograde orbit - with respect to the rotation of the accretion disk - if disrupted, will drive a strong loss of angular momentum in the disk and a surge of accretion. In this scenario, all the disrupted material will ultimately be bound to the black hole, which means more mass will be accreted, driving a stronger flare. After this, the AGN will enter a low-state, due to the formation of a cavity in the accretion disk that is replenished on the much slower viscous time-scale. Their predicted lightcurve shape for a retrograde TDE is remarkably similar to the optical lightcurve of \css.
This scenario requires a population of stellar objects embedded in the disk, which have to dynamically interact to scatter a star close enough to the SMBH to be disrupted \citep{Fabj2020}. The different drag experienced by different objects favours the creation of binaries \citep[e.g.,][]{stone2017} and thus the 2- or 3-bodies scattering necessary for the TDE to happen. If such interactions involve neutron stars or stellar-mass BHs, their coalescence can emit detectable gravitational waves (the AGN channel for mergers, see for example \citealt{grobner2020}).
The rise in X-rays can support this scenario. If the TDE lies in the plane of the disk, especially on a retrograde orbit, the disrupted material will experience many shocks with itself (during the debris circularisation process) and with the material in the disk. These shocks can give rise to X-ray emission, which could then naturally be higher after the flare than before. The short-term variability observed in the X-ray data could then reflect the cahotic process of these shocks.

We can perform an order-of-magnitude calculation, to assess if the TDE scenario is a plausible explanation. If the drop in luminosity is caused by the depletion of the disk, the low luminosity state will continue until the disk will replenish itself, something that happens on the viscous time-scale (see \citealt{stern2018}):
\begin{equation}
\label{tnu}
t_\nu \approx 400 \left( \frac{h/R}{0.05} \right)^{-2}\left( \frac{\alpha}{0.03} \right)^{-1}  \left( \frac{M}{10^8 M_\odot} \right) \left( \frac{R}{150 R_g} \right)^{3/2} \; {\rm years}.
\end{equation}
Where $\alpha$ is the viscosity parameter for which we assume $\alpha=0.3$ for a fully ionised disk \citep{martin19,king07}, $h/R$ is the aspect ratio of the disk (i.e., a measure of the disk thickness) and $R_g$ is the gravitational radius.\\
By considering the time the disk takes to replenish itself, we can infer the value of $R$, the point of impact of the tidal stream, as the disk will drain from that point inwards. The first epoch at which the CSS lightcurve went below the average pre-outburst V band magnitude is around MJD 55685, while the last ZTF photometric observation is at MJD 59231, giving a time interval of roughly 10 years, which gives $R\simeq280~R_g$ for $h/R=0.05$.

The tidal radius, i.e. the radius inside which a star ($\star$) can be tidally disrupted by a BH, is $R_t=R_\star\left({M_{BH}}/{M_\star}\right)^{1/3}$ \citep{rees88,Phinney1989}. For a sun-like star and a $10^7\,M_\odot$ BH, this is $R_t\simeq43~R_g$, much smaller than the inferred point of impact. If the TDE involved a giant star, the tidal radius would be much bigger, enough to make this scenario possible. Massive stars in the proximity of an AGN disk are to be expected \citep{cantiello2021}. On top of this, the aforementioned formula for the tidal radius is derived in the Newtonian limit, which may be inaccurate for a BH with mass $\gtrsim10^7M_\odot$, and the eventual spin of the BH may increase the distance at which stellar disruption is possible \citep{kesden12}. Finally, a thicker disk would also change this estimate.

In the case of the TDE in the disk plane, the limits may be less important, as both the star during the infall towards the SMBH and the TDE ejecta may interact with the disk at larger distances than $R_t$ and ``consume'' a larger portion of the disk. 
Since the object has yet to return to its pre-outburst luminosity, the implied recovery timescale is $>$10 years. At this point the disrupted material should have interacted with the disk out to almost 300$\,R_g$, which would encompass around 0.3$\,\rm M_\odot$ 
for a density of $\rm 10^{-10}~g\,cm^{-3}$ (see \citealt{mckernan2021}, eq. 22). Most of the ejecta will lie in the mid-plane, so only a small portion of the material will be left above/below the disk, which will at later times sink back into the mid-plane as a weak disk.

As detailed in Sec.~\ref{sec:UVOptphot}, we consider several survey data to establish that the low emission state continued at least until January 2021. Unfortunately, the lack of continuous photometric coverage means that we cannot be certain of the duration of the low-emission state, be it either longer or shorter than our estimate. Changing the duration of this state (and thus our estimate of $t_\nu$) would strongly affect the estimate of $R$, the distance of the point of impact.

The absolute magnitude of the event was $M_V=-23$, brighter than what observed in TDEs \citep{vanvelzen20review}. This becomes less concerning if the BH is rapidly spinning: \citet{Leloudas2016}, in fact, interpret the bright transient ASASSN-15lh, with an absolute magnitude at peak of $M=-23.5$, as a TDE around a maximally spinning SMBH with mass above $10^8\,M_\odot$. This is because for a spinning BH, the accretion efficiency can increase up to $\epsilon=0.42$ \citep{bardeen72}, for a prograde orbit. On top of  this, due to the already present accretion disk, the amount of material that can be accreted after the disruption may be higher than in the case of a TDE around a dormant SMBH, naturally leading to higher peak luminosities. An increased accretion efficiency would also justify the relatively high amount of energy radiated during the flare $\rm (\sim 10^{52}\, erg)$. We stress that the direction of the orbit (pro/retrograde) considered for the value of the accretion efficiency is not the same as the direction of the orbit considered in \citet{mckernan2021}: the first refers to the sense of rotation of the accreting material with respect to the spin of the BH; the second to the orbit of the to-be-disrupted star with respect to the rotation of the accretion disk.

The decay rate is different than the canonical $t^{-5/3}$ behaviour of TDEs, but there are many TDEs that deviate from such behaviour \citep{vanvelzen20review} and, moreover, the presence of the AGN accretion disk may well influence the fallback of the material and therefore the decay rate.

Finally, the stacked X-ray spectrum is well fit by a powerlaw with photon index $\rm\Gamma\sim3.3$,  which is in line with what observed in TDEs \citep[see][for a review]{saxton2020}. This soft powerlaw is present in all \textit{Swift}-XRT observations with enough counts for the spectrum to be fit, albeit the photon index is very poorly constrained. Indeed, the \textit{Swift}-XRT observations do not show counts above 5 keV.

Support for a TDE interpretation of \css\ comes from comparing it with other flaring AGNs with similar characteristics. Firstly, we can compare with 1ES 1927+65, a changing look AGN that showed a dramatic brightening in the UV/Optical \citep{trakhtenbrot2019} and a dip in X-ray emission \citep{Ricci2020}. Unlike \css, this AGN also showed dramatic changes in its optical spectra, with the appearance of broad Balmer lines. In \citet{trakhtenbrot2019}, the optical flare is attributed to a sudden increase in accretion rate, with the timescales being difficult to reconcile with standard accretion disk models, while being more in line with a TDE. In \citet{Ricci2020}, the X-ray properties of 1ES 1927+65 are analysed: after the optical outburst, the X-ray luminosity decreased, with the disappearance of the hard power-law component attributed to the destruction of the X-ray corona during the optical flare. The X-ray emission then increases back to the previous levels, and the power law component reappears, hinting at the recreation of the corona. Interestingly, the source showed strong intra-day X-ray variability, similarly to \css. Overall, the X-ray and UV/Optical properties of 1ES 1927+65 have been found more compatible with a TDE scenario, in which the stellar debris disrupt the accretion disk, causing also the destruction of the corona, which is then reformed when the disk is replenished.

Another case that makes for a relevant comparison is PS16dtm, an UV/optical flare happening in a NLS1 galaxy that \citealt{blanchard2017} attributed to a TDE. The galaxy showed an ample UV/optical flare and a dramatic decrease in X-ray emission, compared to archival observations. During this outburst, the emission lines typically found in a NLS1 were seen being enhanced. The authors conclude that a TDE is the most likely scenario to explain all the observed photometric and spectral properties, with the tidal debris obscuring the X-ray emission and perturbing the X-ray corona. They also argue that the similar characteristics between PS16dtm and \css\ (NLS1 host galaxy, three-component emission line structure, peak luminosity at the Eddington level) would favour a TDE scenario to explain the \css\ flare.

\subsubsection*{Rates}
Ultimately, the rates of TDEs of different origins can provide stringent constraints on the dynamics in galactic nuclei. However,
determining the TDE rate is a long-standing challenge. Observationally the detection of TDEs is difficult against the high contrast of galactic nuclei, such that estimating the recovery as a function of magnitude is extremely challenging, and variable on a galaxy to galaxy basis. Furthermore, even when identified, the 
classification of events as TDEs is difficult since nuclear supernovae (of various types) and AGN activity are all confusing sources. 

This may in part explain the order of magnitude difference between the apparently observed ($\rm \dot{N}\simeq10^{-5}\,yr^{-1}gal^{-1}$) and predicted
($\rm \dot{N}\gtrsim 10^{-4}yr^{-1}gal^{-1}$) TDE rates \citep[see][for a review]{stone2020rateschapter}. However, those theoretical estimates are also subject 
to significant uncertainty regarding the stellar populations and density in the nuclei, and indeed
do not consider the impact of the presence
of an accretion disc.
\citet{macleod2020} explores the effect of the presence of an accretion disk on the TDE rate, finding that, while the most eccentric stellar orbits are modified by the accretion disk, the overall TDE rate is mildly affected only for disks with high surface density (like those accreting at near-Eddington rates). More interesting for our case is the fact that repeated disk-crossings can cause the orbit of a star to sink in the accretion disk, and that this effect is stronger for retrograde orbiters. This means that over time the number of stars embedded in the disk may increase, introducing a dependency on the AGN age, regarding the in-disk TDE rate. This dependency will come in the form of a balancing between replenishing mechanisms (such as the one just described) and other phenomena that remove stellar objects from the disk (e.g., stellar inspiraling, scatterings out of the disk).\\

As reported by \citet{mckernan2021}, the most effective way to have a TDE in the disk plane is by 2+1 scatterings (i.e., a stellar object scattered by the interaction with a binary system). This means that the rate of retrograde, in-disk TDEs will depend on the number of objects embedded in the disk and, again, on the AGN lifetime, as objects embedded in the disk tend to merge due to differential drag, reducing the number of scattering binaries in the disk. 
In \citet{mckernan2021}, the rates of in disk TDE (pro or retrograde) are expressed as dependent on the number of embedded objects in the disk and the typical timescale at which star orbits decay towards the inner disk. Given the reasonings above, an additional dependency on the AGN age and on the rate of these mergers may be important, but understanding the precise shape of this dependency is beyond the scope of this paper.

From an observational point of view, only a handful of candidates out of the $\lesssim100$ known TDEs were discovered in an AGN and there is a well know observational bias against TDEs happening in AGNs, which may be mistaken for AGN variability \citep[e.g.,][]{zabludoff2021}. Indeed, the biggest TDE sample from a single survey to date \citep{hammerstein2022} excludes broad-line AGNs from their transient search. Of the 19 TDE host galaxies examined by \citet{hammerstein2021}, only 2 have (narrow) emission line ratios compatible with a pure AGN ionising source (also in this work, broad-line AGNs were excluded from the selection).

\subsubsection{AGN activity}
Since the flare was spatially coincident with the nucleus of \css\ \citep{drake11}, it is straightforward to invoke some form of extreme variability that caused enhanced accretion to explain the observed flare. Even the TDE scenario described in the previous section can be considered an episode of enhanced accretion, and indeed TDEs have been proposed to explain some extreme flares in AGNs \citep[e.g,][]{merloni15}. 

As mentioned previously, the sample of \textit{hypervariable} AGNs is growing in number. For example, \citet{rumbaugh18} present a search for such high amplitude variability in a large sample of quasars, finding that up to 30-50\% of quasars may show such behaviour. There is still no agreement on what could be the driver of such accretion variability as the timescale and intensity of the observed flares are not consistent with standard accretion disk models.
While flares with similar characteristics to the one of \css\ have been observed occurring in AGNs, what is novel, to the best our knowledge, is the subsequent low emission state. This and the anti-correlated behaviour of the X-ray emission are the most complicated aspects to explain in the context of accretion disk variability. 

For the accretion disk to be the source of the observed variability, a relatively large part of the disk would need to be quickly accreted onto the SMBH. This is already problematic per se, as there is no clear mechanism that could cause such a large quantity of angular momentum to be dispersed on the short timescales observed (see the discussion in \citealt{cannizzaro20}). On top of this, in an AGN X-rays are emitted in the innermost part of the accretion disk (soft X-rays) and in a corona (hard X-rays) right above the inner disk itself. It would be natural to expect the X-ray emission to be enhanced when material is quickly drained inwards. What we observe is instead a lower X-ray emission (compared to post-flare data) and the absence of hard X-ray emission, which continued after the flare. Unfortunately, we do not have pre-flare X-ray data to compare the emission level or to confirm if the soft X-ray spectrum is a characteristic of the AGN - in contrast with the typically harder AGN spectra - or of the flare. We note that some NLS1 show a steeper (softer) X-ray powerlaw spectrum when compared with other Seyfert galaxies \citep{Boller1996}, but this typically is due to a soft excess, rather than an absence of hard X-ray photons.

\citet{fragile2020} have shown how, during a type 1 X-ray burst in a low mass X-ray binary, the Poynting-Robertson (PR) drag (induced by the burst radiation field) causes a surge in accretion rate on a faster timescale than that at which the disk replenishes itself and therefore the inner accretion disk around the neutron star recedes outwards. The disk structure then reverts back to its initial condition on the timescale of the burst. A similar phenomenon could explain the lower X-ray emission during the \css\ flare, with the inner accretion disk moving outwards during the rise of the flare and causing the lower X-ray emission. In the case of a NS, the PR drag is caused by radiation emitted from the surface of the NS. An SMBH, naturally, has no such surface, therefore it is unclear - without proper numerical modeling - if an enhanced accretion episode can induce the necessary PR drag to cause the inner disk recession. 

To explain the low state, we could invoke a similar mechanism to the one explored in the previous section: after the drainage of a portion of the disk, the disk itself hasn't been replenished yet and this causes the emission to be lower. In this case, the trigger for the disk drainage is not an external factor (i.e. the stream of disrupted star material) but internal to the accretion disk, a mechanism of yet unknown nature.

An AGN-related explanation of the observed flares of \css\ and PS16dtm was already put forward by \citet{moriya2017}, who proposed the interaction between a disk wind (triggered by limit-cycle oscillations) and the BLR as the source of the observed emission. The fading of the X-ray emission can be then explained by obscuration of the central accretion disk by the wind itself (similarly to the TDE scenario, where the tidal debris obscure the X-ray emitting region). The authors do not investigate in detail the long-term dimming of \css, but we can envision that a {\em fine-tuned} disk wind would be able extract a sufficiently large quantity of angular momentum from the disk to cause a long term change in the accretion rate.
\\

Therefore, while the diversity of nuclear variability in galaxies makes interpreting any single event challenging, the overall properties of \css\ are strikingly similar to the expectations of a disk-draining event, following the tidal disruption of a star within a pre-existing AGN.

\subsection{The nature of the host galaxy}

Another puzzling element of this source is the nature of its host galaxy: it appears to be almost point-like, with no extended component discernible upon visual inspection. With the estimated black hole mass of $\rm 10^7\,M_\odot$, one would expect the host galaxy to be massive and more extended, as it is often the case for galaxies hosting AGNs \citep[e.g.,][]{kauffmann2003}. Indeed, a visual inspection of AGN host galaxies at comparable redshift \citep{Kim2021} shows that these are typically extended and with a generally more complex morphology than the host galaxy of \css. Indeed, of the 154 host galaxies studied with {\em HST} by \cite{Kim2021}, all are readily recoverable in the images via simple visual inspection. Although the objects in their sample all have $z<0.1$, lower than that of CSS100217, the implication is that AGNs in such small host galaxies are extremely unusual. 

Our current mass estimate for the stellar mass of the host galaxy (Sec.~\ref{sec:galfit}) is $M_{\star,gal} < 3\times10^{10}\,M_\odot$. Almost no AGN at this redshift resides in a galaxy with a stellar mass $\lesssim10^{10}\,M_\odot$ \citep{kauffmann2003} and our mass estimate is at the limit to consider this a dwarf galaxy, a type of systems that have been seen showing AGN emission \citep{kaviraj2019}, but with a lower bolometric luminosity ($\rm \lesssim10^{40}$\unitlum) and harbouring BHs with lower masses than what we see in \css\ \citep[e.g.,][]{Mezcua2020}, two properties that appear in support of the dichotomy between the observed AGN and host galaxy of \css. A further constraint is the absence of any stellar absorption lines in high signal to noise spectroscopy of the source. Although the strongest of such lines from the interstellar medium reside in the UV, the absence of, for example, Calcium H\&K absorption is surprising. 

If the host of \css\ is an ultra-compact dwarf then it is relevant to consider its origin. In general such dwarfs are thought to be the stripped cores of larger galaxies. However, there are no obvious galaxies proximate on the sky which are likely of comparable redshift, or show signs of recent tidal interaction. 

The faint host galaxy may, however, have played an important role in identifying a key features of \css, namely the lower-state entered after the outburst. A galaxy in which the light post-outburst was dominated by stellar light would likely have precluded the identification of the lower state since it dilutes the AGN-light. 
In a galaxy with stronger stellar light, identifying the lower state may be more difficult, as the AGN emission would be more diluted, i.e., the contrast between AGN and host galaxy is smaller. In the case of \css, identifying the lower state must mean that the AGN dominates the V band observations (pre- and post-peak). Employing difference imaging, as the most recent wide field surveys do, can counterbalance this effect.

In \citet{Stone2016} it is shown that the TDE rate is enhanced in galaxies with a steeper inner Nuker profile (i.e. with a denser core). Unfortunately, due to the point-like nature of the host galaxy, we were unable to properly model the galaxy to determine the inner Nuker profile. Qualitatively, a compact host galaxy would suggest an enhanced TDE rate, due to the higher stellar densities close to the nucleus - although this could be counter-balanced by the relatively low stellar mass we estimated. In any case, the \css\ host galaxy seems to be more centrally concentrated than a ``regular'' galaxy with a similar mass would be, according to the mass-size relation. Indeed, \citet{stonevanvelzen2016} find an enhanced TDE rate in the centrally overdense galaxy NGC 3156. Moreover, a study of four TDE host galaxies with HST images finds them characterised by a high central stellar density \citep{French2020}. A similar result is found by \citet{graur2018}, using SDSS data. A high TDE rate would naturally favour the TDE scenario.

\section{conclusions}

We discussed the nature of \css, a nuclear transient whose interpretation is debated between a nuclear SN, a TDE or extreme AGN variability. The object showed a strong flare in the optical/UV, with a quick rise to peak and relatively longer decay. The X-ray emission seems to be instead low during the optical outburst and increasing at later times, with a (poorly constrained) soft spectrum at all epochs. The novel aspect is the behaviour of the optical lightcurve of \css, which shows a long-term low luminosity state coincident with the decay of the flare, that is straightforward to attribute to changes in the accreting SMBH and therefore seemingly excludes a SN as a possible interpretation of the event. Comparison with other similar objects and recent theoretical work \citep{chan19,mckernan2021} support a TDE interpretation of the flare, where the stream of disrupted material perturbes the accretion disk creating an empty cavity that has yet to be replenished and the X-ray emission is shielded from sight by the disrupted material. We cannot exclude that the flare and the low luminosity state are instead caused by an unknown form of AGN variability.

Continuous monitoring of the source will be important to assess if and when the luminosity of \css\ will reach the pre-outburst level. Knowing the time at which the disk will be replenished will allow for a determination of the point of impact of the tidal stream, or for the exclusion of the TDE scenario. Alternatively, observing a second flare would point at AGN variability as the source of the observed emission. Additionally, with further, high-quality X-ray data, the slope of the spectrum can be better determined at late times, which could help discern between an AGN flare and a TDE.

\css, as a SMBH accretion flare of either AGN or TDE origin, joins the ever growing ranks of rapid and intense nuclear flares happening in NLS1 galaxies, such as PS16dtm \citep{blanchard2017}, AT2017bgt \citep{trakhtenbrot2019}, F01004-2237 \citep{tadhunter17,trakhtenbrot2019,cannizzaro2021} and, more recently, the 5 flares presented in \citet{frederick2021}. 
The BH mass estimated for \css\ is low enough to fit in one of the scenarios proposed by \citet{frederick2021} to explain this apparent host galaxy preference: a selection effect due to the shorter timescales connected to a lower BH mass. As some of the \citet{frederick2021} flares are attributed to a TDE, we checked the latest ZTF data\footnote{\url{https://irsa.ipac.caltech.edu/frontpage/}} to check for a low-state after the flare, but found that none of the sources show such behaviour.

\css\ shows a strong IR echo, probably emitted by dust in the system. This is a characteristic shared by many nuclear flares (e.g., AT2017gbl, a flare in an AGN interpreted as a TDE by \citealt{kool20}) and was at the base of the search for neutrino-emitting flares of \citet{vanvelzen2021neutrino}, where super Eddington accretion is proposed as the mechanism driving the neutrino emission. We note that \css\ reached the Eddington luminosity at the peak of its V band lightcurve.

\css\ has also another peculiar characteristics: its host galaxy is smaller and less massive than other AGN hosts at comparable redshifts, but the BH mass and AGN bolometric luminosity are higher than what we would expect from an AGN in a dwarf galaxy. Such a compact profile could cause an enhancement in the TDE rates of the galaxy. Unfortunately, due to the nature of our data, we were unable to put strong constraints on the morphology of the host galaxy. In fact, the HST observations of \css\ were made during the outburst and with the scientific goal of investigating the transient, rather than the host galaxy: the exposure times are relatively short and the filters employed not the most apt to study the galaxy (e.g., the redder filter, F763M is a medium-band filter and therefore captures less light). Observations focused at investigating the nature of the host galaxy, now that the source is in a relatively low luminosity state, will allow to cast light on its mass and morphology.

\section*{acknowledgements}
We thank the referee for a very constructive report which helped to improve the quality of this paper. 

GC and AJL have received funding from the European Research Council (ERC) under the European Union’s Seventh Framework Programme (FP7-2007-2013) (Grant agreement No. 725246).

\subsection*{Data Availability}
All data will be made available in a reproduction package uploaded to Zenodo. 

\bibliographystyle{mnras}
\bibliography{bibliography}

\bsp	
\label{lastpage}
\end{document}